\documentstyle[epsf]{elsart}  
\begin{document}
\newcommand{\p}{\partial}
\newcommand{\ls}{\left(}
\newcommand{\rs}{\right)}
%%%%%%%%%%%%%%%%%%%%%%%%%%%%%%%%%%%%%%%%%%%%%%%%%%%%%%%%%%%%%%%%%%%%%%%%%
%                                                                       %
%   BEGIN OF DOCUMENT                                                   %
%                                                                       %
%%%%%%%%%%%%%%%%%%%%%%%%%%%%%%%%%%%%%%%%%%%%%%%%%%%%%%%%%%%%%%%%%%%%%%%%%
\begin{frontmatter}
\title{Relativistic heavy ion collisions with realistic 
non-equilibrium mean fields} 
\author[tuebingen]{C. Fuchs}
\author[muenchen]{T. Gaitanos}
\author[muenchen]{H.~H. Wolter}
\address[tuebingen]{Institut f\"ur Theoretische Physik der 
Universit\"at T\"ubingen, D-72076 T\"ubingen, Germany}
\address[muenchen]{Sektion Physik, Universit\"at M\"unchen, 
D-85748 Garching, Germany}  
%**************************************************************************
\begin{abstract}
We study the influence of non-equilibrium 
phase space effects on the dynamics of heavy ion reactions 
within the relativistic BUU approach.
We use realistic Dirac-Brueckner-Hartree-Fock (DBHF) mean fields 
determined for two-Fermi-ellipsoid configurations, 
i.e. for colliding nuclear matter, in a local phase space 
configuration approximation (LCA). We compare to DBHF 
mean fields in the local density approximation (LDA) 
and to the non-linear Walecka model. The results are further compared 
to flow data of the reaction $Au$ on $Au$ at 400 MeV per 
nucleon measured by the FOPI collaboration. We find that the 
DBHF fields reproduce the experiment if the 
configuration dependence is taken into account. This has also 
implications on the determination of the equation of state from 
heavy ion collisions. 
\end{abstract}
\end{frontmatter}
%%%%%%%%%%%%%%%%%%%%%%%%%%%%%%%%%%%%%%%%%%%%%%%%%%%%%%%%%%%%%%%%%%%%%%%%%
%                                                                       %
%   BEGIN OF TEXT                                                       %
%                                                                       %
%%%%%%%%%%%%%%%%%%%%%%%%%%%%%%%%%%%%%%%%%%%%%%%%%%%%%%%%%%%%%%%%%%%%%%%%%
\section{Introduction}
%%%%%%%%%%%%%%%%%%%%%%%%%%%%%%%%%%%%%%%%%%%%%%%%%%%%%%%%%%%%%%%%%%%%%%%%
A principal object of the investigation of heavy ion collisions 
at intermediate energies is to determine the nuclear equation of state (EOS), 
i.e. the properties of nuclear matter in equilibrium away from saturation and at 
non-zero temperatures. However, the phase space distribution in a heavy 
ion collision is out of global and even local equilibrium through 
most of the collision time. Thus transport models have been 
developed to describe the evolution of the phase space 
\cite{bg88,aist86,koli88,bkm93,fu95}. 
Although these models are quite successful to reproduce data 
the attempts to determine the EOS, e.g., from the transverse flow 
in heavy ion collisions have not led to generally accepted results. 
Recently also experimental collaborations, like FOPI or EOS, again focused on this 
problem \cite{fopi1,eos95}. 

The difficulty is in part due to the non-equilibrium effects mentioned above. 
The common practice of theoretical 
calculations is to apply phenomenological mean fields 
like Skyrme forces \cite{aist86} in a non-relativistic 
or the Walecka model \cite{wal74} and its non-linear 
extensions in a relativistic approach \cite{bkm93,Boguta77}. 
These forces contain parameters which allow 
to vary the characteristics of the corresponding EOS. 
In some cases a phenomenological momentum dependence taken 
from the empirical nucleon-nucleus optical potential has been 
added \cite{aist86,mbc94}. 
However, non-equilibrium features of the phase space are 
not contained in these forces, a fact which certainly will lead 
to uncertainties when conclusions on the ground state EOS are drawn.

A derivation of a kinetic equation 
from a microscopic non-equilibrium many-body theory 
shows that a consistent treatment requires to determine 
the effective interaction in the nuclear 
medium for the non-equilibrium phase space configurations of 
heavy ion collisions \cite{btm90}. Generally, the 
Dirac-Brueckner-Hartree-Fock (DBHF) theory provides a very successful 
approach to the many-body problem in nuclear matter \cite{hs87,thm87a} 
and in finite nuclei \cite{bt92,Boema94,fule95}. 
However, a solution of the Bethe-Salpeter equation for arbitrary 
anisotropic momentum configurations is extremely difficult. 
Approximate treatments should retain the most important features of such 
configurations. To improve on the local 
density approximation (LDA) the phase space has been represented locally 
by two Fermi ellipsoids separated by a relative momentum. We have called this the 
{\em local configuration approximation} (LCA) \cite{fu92}. In contrast 
to the LDA which refers to ground state nuclear matter the LCA describes 
colliding nuclear matter (CNM) and should be able to reproduce 
the main features of the anisotropic momentum space in a heavy ion 
collision over the entire reaction time.

However, even for the idealized momentum configuration of
two Fermi ellipsoids the Bethe-Salpeter equation 
has not yet been solved. Only in a non-relativistic 
framework this problem has been treated before \cite{jae92}. 
Therefore as a first step Sehn et al. \cite{sehn90,sehn95} 
developed a procedure to construct 
self-energies for colliding nuclear matter from an appropriate 
parametrization of DBHF ground state results. These self-energies 
take into account specific non-equilibrium features of the 
momentum space and approximately include the correlations 
of the $T$-matrix. Rather than introducing model 
parameters as, e.g. in the Skyrme or the 
Walecka model, the DBHF mean fields are connected in a parameterfree 
way to the free NN-interaction. In this sense we call these fields 
"realistic".

We compare the present approach to experiment, i.e. to 
FOPI flow data of the system $Au$ on $Au$ at 400 A.MeV 
\cite{fopi1,gobbi94}. In addition we compare 
to the LDA and to a standard force used in relativistic heavy ion 
calculations, the non-linear Walecka model (NL2). 
%%%%%%%%%%%%%%%%%%%%%%%%%%%%%%%%%%%%%%%%%%%%%%%%%%%%%%%%%%%%%%%%%%%%%%%%%%%%%%%
\section{DBHF Mean Fields in the RBUU Approach}
%%%%%%%%%%%%%%%%%%%%%%%%%%%%%%%%%%%%%%%%%%%%%%%%%%%%%%%%%%%%%%%%%%%%%%%%%%%%%%%
As shown, e.g., in Ref. \cite{btm90} the self-energy 
$\Sigma = \Sigma_{s} - \gamma^\mu \Sigma_\mu $ 
which enters into the relativistic Boltzmann-Uehling-Uhlenbeck (RBUU) 
equation  
\begin{equation}
 \left[p^{*\mu} \partial_{\mu}^x  + \left( p^{*}_{\nu} F^{\mu\nu}     
+ m^* \partial^{\mu}_x m^* \right) 
\partial^{p^*}_{\mu} \right] f (x,p^* ) = I_C
\label{TP}
\end{equation}
via effective masses $m^* ={\mathrm{M}} +  {\mathrm{Re}}\Sigma_s $, 
kinetic momenta 
$p^{*}_\mu = p_\mu + {\mathrm{Re}}\Sigma_\mu $, and 
the field strength tensor $F^{\mu\nu}$ 
in principle has to be determined from the corresponding $T$-matrix 
in non-equilibrium. Consistently with the kinetic 
equation (\ref{TP}) we represent $\Sigma$ in Hartree form by seperating 
off a linear density dependence through the 
definition of dynamical coupling functions  
\begin{equation}
Re \Sigma [f]  = - \Gamma_{s} [f]  \rho_{s}  [f] 
          + \gamma_\mu \Gamma_0 [f] j^\mu  [f]  
\quad .
\label{sigma_12_H}
\end{equation}
In Eq. (\ref{sigma_12_H}) $\rho_{s}$ and $j^\mu$ are the scalar density 
and the baryonic four-current, repsectively, and $[f]$ denotes the dependence 
on the phase space distribution $f$. The self-energy, Eq. (\ref{sigma_12_H}), 
is of the same structure as the mean field 
in the $\sigma\omega$--model \cite{wal74}, however, the 
coupling constants $\frac{g_\sigma^2}{m_\sigma^2}, \frac{g_\omega^2}{m_\omega^2} $ 
for the scalar ($\sigma$) and the vector 
($\omega$) meson are replaced by the dynamical 
coupling functions $\Gamma_{s,0}[f]$. These are Lorentz scalars 
and given by invariants of the $T$-matrix averaged over $f$ \cite{sehn95}. 
In the LCA the $\Gamma_{s,0}$, Eq. (\ref{sigma_12_H}), are locally 
approximated by the values for the corresponding colliding 
nuclear matter configurations, and thus depend on the 
collective parameters of the CNM configuration, i.e. the 
Fermi momenta and the relative velocity of 
the local currents 
\begin{equation}
\Gamma_{s,0}  [f]  \longmapsto 
{\overline \Gamma}^{(12)}_{s,0} 
(p_{F_1},p_{F_2},v_{\mathrm{rel}})   
\quad .
\label{coupl}
\end{equation}
In this formulation the LDA means 
to use ${\overline \Gamma}(p_{F_{\mathrm{tot}}})$ corresponding to 
a single Fermi sphere at the respective total density. 
The calculation of the ${\overline \Gamma}^{(12)}_{s,0}$ is 
discussed in detail in Ref. \cite{sehn95}. They are constructed 
by extrapolating the density and momentum dependent coupling functions 
$\Gamma_{s,0}(p,\rho_B)$ defined as in Eq. (\ref{sigma_12_H}) 
for DBHF nuclear matter calculations to CNM configurations. There 
we used the DBHF results of Ref. \cite{thm87a} (without $\Delta$-particles). 
Thus the T=0 EOS is the one given in Fig. 3.8 of Ref. \cite{thm87a}, 
which gives a good fit of the saturation properties, an incompressibility 
of 250 MeV, and a good reproduction of the energy dependence of 
real part of the optical potential up to about 400 MeV.   

In the LCA the collective parameters of the CNM configurations 
are determined in every time step from the actual 
phase space distribution. To do so $f$ is decomposed into 
the respective contributions from projectile 
and target $ f=f^{(1)} + f^{(2)}$.  
The Fermi momenta 
$p_{F_i} = (2/3\pi^2 \rho_{0}^{(i)})^{\frac{1}{3}}$ are defined 
in the rest frames of the currents by an invariant 
rest density \cite{fu92,sehn90}
\begin{eqnarray}
\rho_{0}^{(i)} (x) &=& \sqrt{ j_{\mu}^{(i)}  (x) j^{ (i)\mu}(x) }
\quad ,\quad i=1,2
\label{rho_LCA}
\\
j_{\mu}^{(i)} (x) &=& 4\int \frac{d^3 p}{(2\pi)^3} 
\frac{ p_{\mu}^*}{E^*} f^{(i)} (x,{\vec p})  
 = u_{\mu}^{(i)} (x) \rho_{0}^{(i)} (x)
\quad .
\label{current_LCA}
\end{eqnarray}
The invariant relative velocity 
$v_{\mathrm{rel}}=|{\vec v}_{\mathrm{rel}} |$ is obtained 
from the streaming velocities 
$u_{\mu}^{(i)} = (u_{0}^{(i)},{\vec u}^{(i)})$ as
\begin{equation}
{\vec v}_{\mathrm{rel}} (x) = \frac{u_{0}^{(2)} {\vec u}^{(1)} - u_{0}^{(1)} 
{\vec u}^{(2)} }  {u_{\mu}^{(1)} u^{(2)\mu} }
\label{vrel_LCA}
\quad .
\end{equation}
In the present calculations we assume symmetric configurations, 
i.e. $p_{F_1}=p_{F_2}$, 
which is a good approximation to the participant 
region. For highly asymmetric configurations where one density falls 
significantly below the other, what mainly occurs in the more 
peripheral reactions, the system is treated as one Fermi sphere in the LDA.  

\begin{figure}[b]
\begin{center}
\leavevmode
\epsfxsize = 10cm
\epsffile[30 85 430 410]{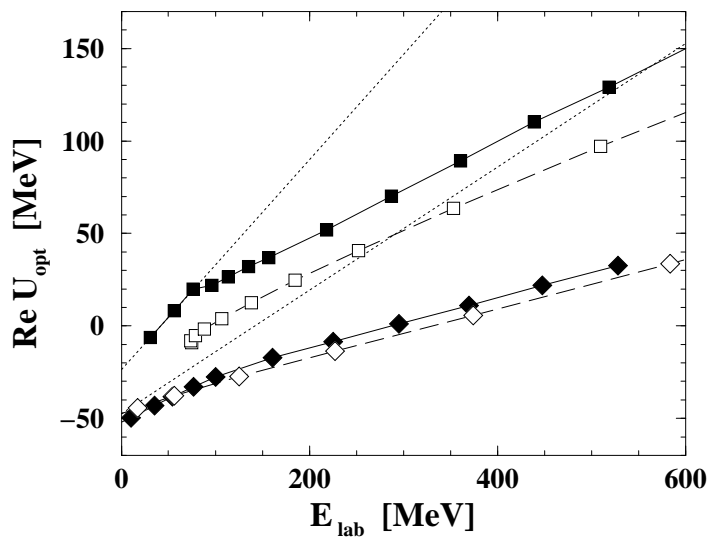}
\end{center}
\caption{\label{opt_graph}
Energy dependence of the DBHF optical potential. The solid lines 
represent the nucleon--nucleus optical potential taken from ref. 
\protect\cite{thm87a} at saturation density $\rho_{\mathrm{sat}}$ (diamonds) 
and 2$\rho_{\mathrm{sat}}$ (squares). The dashed lines represent the 
corresponding nucleon optical potential in a nucleus--nucleus 
collision determined in the colliding nuclear matter 
approximation at subsystem densities 
$\rho_{0}^{(1)}+\rho_{0}^{(2)} =\rho_{\mathrm{sat}}$ (diamonds) and 
$\rho_{0}^{(1)}+\rho_{0}^{(2)} =2 \rho_{\mathrm{sat}}$ (squares). The 
dotted lines refer to the DBHF nucleon--nucleus optical 
potential obtained in a simple LDA with no momentum 
dependence included at densities $\rho_{\mathrm{sat}}$ (lower curve) 
and 2$\rho_{\mathrm{sat}}$ (upper curve).
}
\end{figure}

To give an impression of the configuration effects 
we show in Fig. \ref{opt_graph} the Schr\"odinger 
equivalent real part of the 
optical potential as a function of $E_{\mathrm{lab}}$. The DBHF 
nucleon optical potential in a nucleus--nucleus collision 
in the CNM approximation \cite{sehn95} is compared to the corresponding 
nucleon--nucleus optical potential \cite{thm87a} at the same density. 
In CNM the energy dependence of $U_{\mathrm{opt}}$ 
originates from the dependence 
of the mean field on $v_{\mathrm{rel}} (E_{\mathrm{lab}})$ whereas in the 
LDA, also shown in Fig. \ref{opt_graph}, 
the self-energy is taken at the corresponding density and kept constant. 
In the latter case $\mathrm{Re}U_{\mathrm{opt}}$ scales linearly with energy as 
is the case, e.g., also in the Walecka model \cite{bkm93,wal74}. 
The configuration dependence (LCA) significantly softens the optical 
potential leading to a similar 
behavior as in nuclear matter when the full momentum dependence 
of the self-energy is included \cite{thm87a}. At two times 
saturation density the nucleus--nucleus optical potential is even softer. 
%%%%%%%%%%%%%%%%%%%%%%%%%%%%%%%%%%%%%%%%%%%%%%%%%%%%%%%%%%%%%%%%%%%%%%%
%%%%%%%%%%%%%%%%%%%%%%%%%%%%%%%%%%%%%%%%%%%%%%%%%%%%%%%%%%%%%%%%%%%%%%%
More exactly speaking we use "cold" CNM configurations to 
parametrize the phase space distribution, i.e. two sharp 
Fermi spheres. More consistently one could use two Fermi spheres 
of finite temperature. In a forthcoming work we analysed 
the phase space distribution in such a way and we find 
temperatures of less than about 30 MeV. In Ref. \cite{thm87a} 
it is found that $\mathrm{Re}U_{\mathrm{opt}}$ is only weakly changed at these 
temperatures and thus the assumption of zero temperature fields 
appears justified.

For the solution of the relativistic BUU equation we use the 
relativistic Landau-Vlasov method \cite{fu95} which represents the 
phase space distribution by covariant gaussian testparticles 
in coordinate and momentum space. 
A fully consistent treatment of the kinetic equation in principle 
requires to treat the in-medium cross section on 
the same footing as the mean field \cite{btm90}. 
Hence in Ref. \cite{fu96} 
an in-medium cross section has been derived for CNM 
from the imaginary part of the DBHF self-energies. 
However, in this work we focus on the mean field and apply 
the standard Cugnon parametrization \cite{cug81} for the cross section. 
As shown, e.g., in Ref. \cite{khoa92} the transverse flow does not react 
very sensitively to the cross section and we do not expect the present 
results to be significantly altered by the use of a 
consitently determined in-medium cross section.  
%%%%%%%%%%%%%%%%%%%%%%%%%%%%%%%%%%%%%%%%%%%%%%%%%%%%%%%%%%%%
\section{The System Au on Au at 400 A.MeV}
%%%%%%%%%%%%%%%%%%%%%%%%%%%%%%%%%%%%%%%%%%%%%%%%%%%%%%%%%%%%
The present system is particularly well adapted to study 
non-equilibrium effects. The bombarding energy is high enough to 
result in distinct two-Fermi-ellipsoid momentum configurations 
in the initial phase of the reaction and the densities reached in the 
compression phase are sufficiently high to test the density dependence 
of the EOS. Impact parameter selected high quality data 
of nuclear flow are provided by the FOPI collaboration \cite{fopi1}. 
%%%%%%%%%%%%%%%%%%%%%%%%%%%%%%%%%%%%%%%%%%%%%%%%%%%%%%%%%%%%%%%%%%%%%%%
\begin{figure}[h]
\begin{center}
\leavevmode
\epsfxsize = 12cm
\epsffile[0 190 600 600]{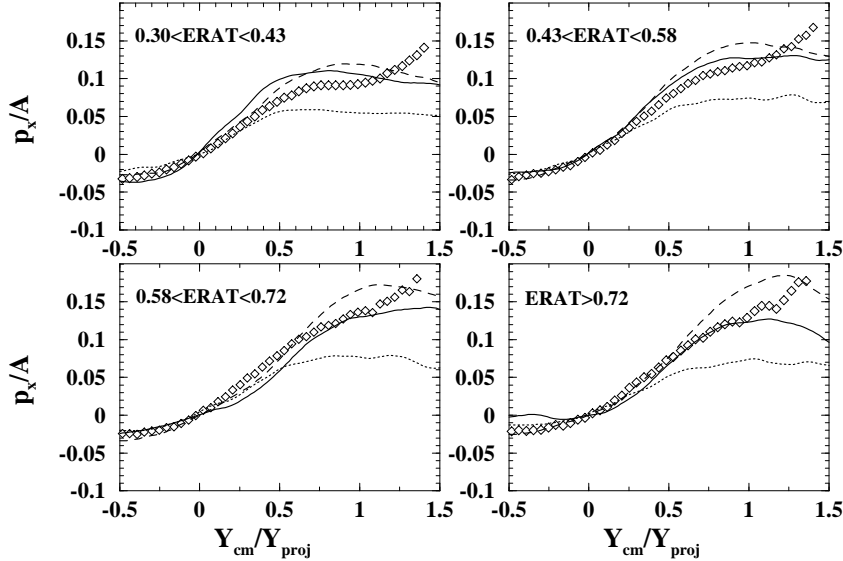}
\end{center}
\caption{\label{flow_graph}
Transverse flow per nucleon in units of the nucleon mass for 
the reaction $Au$ on $Au$ at 400 A.MeV. Four $ERAT$ bins running 
from semi-peripheral (top left) to central collisions (bottom right) 
collisions have been selected. Non-equilibrium 
DBHF mean fields were used (solid curves) as well as DBHF    
mean fields in the LDA (dashed curves). The dotted curves 
correspond to the non-linear Walecka model NL2 and the 
diamonds represent the FOPI data of Ref. \protect\cite{fopi1}. 
}
\end{figure}
%%%%%%%%%%%%%%%%%%%%%%%%%%%%%%%%%%%%%%%%%%%%%%%%%%%%%%%%%%%%%%%%%%%%%%%
Fig. 2 shows the in-plane transverse flow per 
particle $p_{x} /A$ in units of the 
nucleon mass for four different $ERAT$ bins. $ERAT$ is defined as 
the ratio of transversal to longitudinal kinetic energy deposited 
in the forward center-of-mass hemisphere \cite{fopi1} and provides 
a measure for the impact parameter bins 
which run from semi-peripheral ($b\simeq 6$ fm, 
$0.30 \leq ERAT\leq 0.43$) to central ($ ERAT\geq 0.72$) collisions. 
For the analysis of the reaction we use the geometrical part of the 
FOPI filter \cite{reisdorf} which takes into account acceptance cuts 
for different mass fragments. We therefore generated fragments 
after the collision (60 fm/c) by a phase space coalescence model. 
The influence of the filter on the observables 
is similar as in Ref. \cite{fopi1}. For the original experimental 
data the crossover of the flow from negative to positive values 
is slightly shifted to positive center-of-mass rapidity values. 
This behavior is quantitatively not well understood but is supposed 
to be due to recoil effects or double hits \cite{reisdorf}. 
After applying the filter analysis the crossover of the theoretical 
curves is slightly shifted to negative rapidities. 
For a better comparison of the flow, 
i.e. the slope of $p_{x} /A$, in Fig. 2 we readjusted the 
crossovers to $p_{x} /A = 0$ for theory and experiment.

It is seen from Fig. 2 that the non-equilibrium DBHF mean fields (LCA) 
are able to reasonably reproduce the data. 
The configuration dependence of the fields weakens their repulsion, as seen 
in Fig.1, resulting in less flow compared to a pure LDA treatment. 
This configuration effect on the flow is strongly increasing with 
centrality and reaches a magnitude comparable to different 
equations of state, i.e. about 30\% of the total amount of flow. 
This can be understood by the increasing 
overlap of the nuclei since only the participant matter feels 
configuration effects. For the spectator matter our description 
always results in a LDA treatment. We also find that 
over the entire impact parameter range NL2 yields a too small 
transverse flow compared to the experiment. 
%%%%%%%%%%%%%%%%%%%%%%%%%%%%%%%%%%%%%%%%%%%%%%%%%%%%%%%%%%%%%%%%%%%%%%%
\begin{figure}[h]
\begin{center}
\leavevmode
\epsfxsize = 10cm
\epsffile[110 310 450 610]{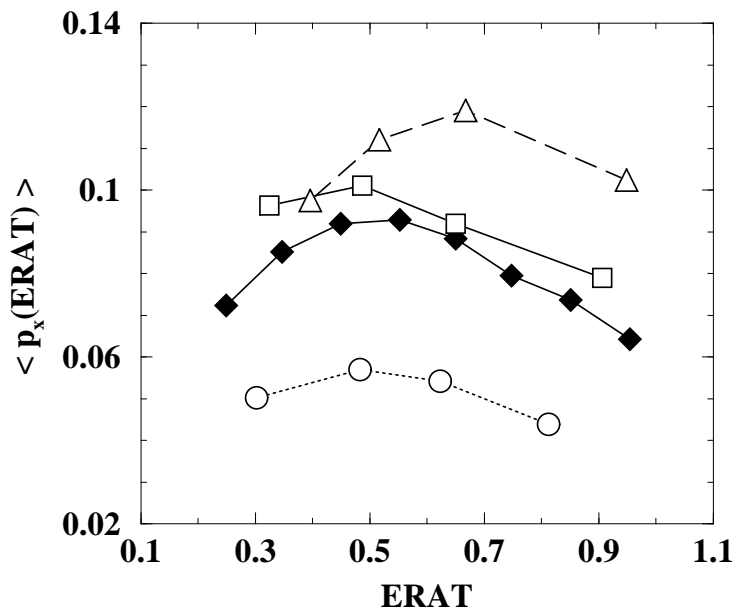}
\end{center}
\caption{\label{px_erat_graph}
Total transverse flow averaged over the foreward cms hemisphere 
in units of the nucleon mass for the same reaction as in Fig.2. 
Non-equilibrium DBHF mean fields were used (squares) as well as DBHF    
mean fields in the LDA (triangles). The circles 
correspond to the non-linear Walecka model NL2 and the 
diamonds represent the FOPI data of Ref. \protect\cite{fopi1}. 
}
\end{figure}
%%%%%%%%%%%%%%%%%%%%%%%%%%%%%%%%%%%%%%%%%%%%%%%%%%%%%%%%%%%%%%%%%%%%%%%%%

In Fig.3 we show the total amount of flow $<p_{x} (ERAT)>$ integrated 
over the forward center-of-mass hemisphere. Again, the LCA calculations are 
close to the data and also reproduce best the shape over the entire 
$ERAT$ range. The larger deviation in the more peripheral 
collisions is probably due to the fact that the CNM mean fields are 
always determined for symmetric configurations 
($p_{F_1}=p_{F_2}$). 
For the very central collisions (high $ERAT$ values) 
the LDA completely fails and overestimates the data by about a 
factor of 2 but becomes more reliable with increasing impact parameter. 
The NL2 model again strongly underestimates the flow. 

The above results are summarized in Table 1. There the mean 
in-plane directed transverse flow normalized to the 
center-of-mass projectile momentum per nucleon 
is given. The experimental value \cite{gobbi94} has been obtained for the PM4 
event class corresponding to semi-central ($3\leq b\leq 5$) collisions. 
This quantity provides a good measure of the global repulsion of the model. 
The DBHF result is close to the experimental value 
when treated in the local configuration approximation (LCA). The local 
density approximation (LDA) overestimates the experiment by about 30\% 
and the NL2 parameter set of the non-linear Walecka model 
underestimates the data by about 40\%.
%%%%%%%%%%%%%%%%%%%%%%%%%%%%%%%%%%%%%%%%%%%%%%%%%%%%%%%%%%%%%%%%%%%%%%%%%
\begin{table}
\caption{\label{satprop}
Mean directed in-plane transverse flow normalized to the center-of-mass 
projectile momentum per nucleon. Calculations were performed with 
DBHF mean fields in non-equilibrium (LCA) as well as in the local density 
approximation (LDA) and with the non-linear Walecka model NL2. The 
experimental value is taken from Ref. \protect\cite{gobbi94}. 
}
\begin{tabular}{ccccc}
\hline\hline \\
                            & \multicolumn{2}{c}{DBHF} &  NL2  & EXP \\  
                            & LCA     & LDA    &        &       \\ \hline
$P_{x}^{\mathrm{dir}}/P_{CM}^{\mathrm{proj}}
$ & 0.187   & 0.232  & 0.103  & 0.173 \\
[0.2ex] \hline\hline
\end{tabular}
\end{table}
%%%%%%%%%%%%%%%%%%%%%%%%%%%%%%%%%%%%%%%%%%%%%%%%%%%%%%%%%%%%%%%%%%%%%%%%%

To understand the influence of the mean field on the reaction dynamics 
we emphasize that in a relativistic approach the 
major part of the repulsion is due 
to Lorentz forces generated by the vector field (see Fig. 1). 
The repulsion and, correspondingly, the stiffness 
of the EOS in first order is given by the effective mass; 
the smaller the effective mass the more repulsive is the 
model \cite{koli88,mbc94}. 
Thus NL2 with a large effective mass ($m^*/m_0$=0.83) 
is relatively soft whereas the DBHF EOS ($m^*/m_0$=0.586) 
is rather stiff. The momentum dependence of the mean field which 
is taken into account in the 
LCA in an averaged form actually weakens this repulsion as 
also found in Ref. \cite{mbc94}. The small 
vector fields of NL2 lead to an 
underestimation of the flow. Furthermore, the density dependence of NL2 
is too small to produce a sufficient amount of flow 
just by compression. Thus the use of parametrizations with a stiffer 
EOS but with the same $m^*$ will not significantly 
improve on this \cite{bkm93}. 

The same reaction as above has been analyzed in the QMD approach in 
Ref. \cite{fopi1} using forces based on the non-relativistic ground state 
BHF $G$-matrix \cite{jae92,khoa92}. There it was found that 
the strong and repulsive momentum dependence of the $G$-matrix 
is necessary to reproduce the data of the more peripheral collisions. 
However, in the more central collisions 
the non-relativistic $G$-matrix lacks a sufficient density 
dependence and results in a too small flow \cite{fopi1}. This 
drawback is resolved in a relativistic approach where the $T$-matrix 
provides the appropriate balance between density and momentum dependence.

%%%%%%%%%%%%%%%%%%%%%%%%%%%%%%%%%%%%%%%%%%%%%%%%%%%%%%%%%%%%%%%%%%%%%%% 
\section{Summary and Conclusions}
%%%%%%%%%%%%%%%%%%%%%%%%%%%%%%%%%%%%%%%%%%%%%%%%%%%%%%%%%%%%%%%%%%%%%%% 
Realistic DBHF mean fields have been used in a detailed 
flow analysis of heavy ion collisions at intermediate energies. 
We have taken into account non-equilibrium features of the phase space 
in a local configuration approximation using mean fields determined for 
a colliding nuclear matter scenario. The ground state DBHF self-energies are 
rather repulsive relative to, e.g., standard parametrizations 
of the non-linear Walecka model. However, 
if the configuration effects are taken 
into account properly the DBHF mean fields are able to reproduce the 
overall behavior of the experimental flow data. Thus, the configuration 
effects turned out to be of major importance, especially if more central 
collisions are considered. A simple LDA treatment 
results in too repulsive mean fields. The intrinsic momentum dependence 
of the local configuration approximation weakens this repulsion 
considerably. An extension of the colliding 
nuclear matter approximation to anisotropic configurations 
and the inclusion of finite temperature effects and of 
an in-medium cross section may still improve the present 
results. A more detailed comparison to data including also 
longitudinal flow variables will be forthcoming.

To summarize we conclude that DBHF mean fields based on realistic NN 
interactions are able to reproduce intermediate energy flow data, thus 
testing the DBHF approach also at high densities. It is found 
that the inclusion of non-equilibrium effects is important in order 
to learn something about the equation of state of ground state 
nuclear matter.\\ 
\begin{ack}
The authors would like to thank L. Sehn 
for valuable discussions. In particular, we thank W. Reisdorf and 
P. Dupieux for helpful discussions with respect to the comparison 
to the data and for providing us with the FOPI filter simulation code. 
\end{ack}
%%%%%%%%%%%%%%%%%%%%%%%%%%%%%%%%%%%%%%%%%%%%%%%%%%%%%%%%%%%%%%%%%%%%%%%%%
%                                                                       %
%   END OF TEXT                                                         %
%                                                                       %
%%%%%%%%%%%%%%%%%%%%%%%%%%%%%%%%%%%%%%%%%%%%%%%%%%%%%%%%%%%%%%%%%%%%%%%%%

%%%%%%%%%%%%%%%%%%%%%%%%%%%%%%%%%%%%%%%%%%%%%%%%%%%%%%%%%%%%%%%%%%%%%%%%%
%                                                                       %
%   END OF THEBILIOGRAPHY                                               %
%                                                                       %
%%%%%%%%%%%%%%%%%%%%%%%%%%%%%%%%%%%%%%%%%%%%%%%%%%%%%%%%%%%%%%%%%%%%%%%%%
\end{document}